%% file: paper.tex
\RequirePackage[switch]{lineno}
\documentclass[twocolumn,showpacs,aps,prl,superscriptaddress,floatfix,letterpaper]{revtex4}

\usepackage{graphicx}
\usepackage{dcolumn}
\usepackage{amsmath}
\usepackage{epsfig}
\usepackage{setspace}

\newcommand{\ifb}{\ensuremath{\mathrm{fb^{-1}}}}
\newcommand{\TeV}{\ensuremath{\mathrm{Te\kern -0.1em V}}}
\newcommand{\TeVc}{\ensuremath{\mathrm{Te\kern -0.1em V\!/}c}}
\newcommand{\TeVcc}{\ensuremath{\mathrm{Te\kern -0.1em V\!/}c^2}}
\newcommand{\GeV}{\ensuremath{\mathrm{Ge\kern -0.1em V}}}
\newcommand{\GeVc}{\ensuremath{\mathrm{Ge\kern -0.1em V\!/}c}}
\newcommand{\GeVcc}{\ensuremath{\mathrm{Ge\kern -0.1em V\!/}c^2}}
\newcommand{\MeV}{\ensuremath{\mathrm{Me\kern -0.1em V}}}
\newcommand{\MeVc}{\ensuremath{\mathrm{Me\kern -0.1em V\!/}c}}
\newcommand{\MeVcc}{\ensuremath{\mathrm{Me\kern -0.1em V\!/}c^2}}

\newcommand{\um}{\ensuremath{\mathrm{\mu m}}}

\newcommand{\babar}{\mbox{\slshape B\kern-0.1em{\small A}\kern-0.1em B\kern-0.1em{\small A\kern-0.2em R}}}

\def\yieldone{19^{+6}_{-5} }
\def\yieldtwo{39\pm11 }

\def\massfitone{4143.5\pm2.8~\MeVcc }

\def\widthfitone{15.2\pm5.9~\MeVcc }

\def\massdifffitone{1046.7^{+2.9}_{-3.0}\pm0.6 }
\def\widthfitone{15.3^{+10.4}_{-6.1}\pm2.5 }

\def\massfitone{4143.4^{+2.9}_{-3.0}~\MeVcc }

\def\massone{4143.4^{+2.9}_{-3.0}(\mathrm{stat})\pm0.6(\mathrm{syst})~\MeVcc }
\def\widthone{15.3^{+10.4}_{-6.1}(\mathrm{stat})\pm2.5(\mathrm{syst})~\MeVcc }
\def\yieldone{19\pm6(\mathrm{stat})\pm3(\mathrm{syst}) }
\def\yieldonetab{19\pm6\pm3 }

\def\rebfone{0.149\pm0.039(\mathrm{stat})\pm0.024(\mathrm{syst}) }

\def\yieldtwo{22\pm8 }

\def\masstwosys{4274.4^{+8.4}_{-6.7}(\mathrm{stat})\pm1.9(\mathrm{syst})~\MeVcc }
\def\widthtwosys{32.3^{+21.9}_{-15.3}(\mathrm{stat})\pm7.6(\mathrm{syst})~\MeVcc }

\begin{document}

\begin{flushleft}
\ \hfill\ \\
\ \hfill\ \\
\ \hfill\ \\
\end{flushleft}

\title
{
{\large \bf 
Observation of the $Y(4140)$ structure in the $J/\psi\,\phi$ Mass Spectrum in $B^\pm\to J/\psi\,\phi K^\pm$ Decays 
} 	
}

\input March2010_Authors.tex

\date{\today}

\begin{abstract}
 
The 
observation of the $Y(4140)$ structure  
in  $B^\pm\rightarrow J/\psi\,\phi K^\pm$ decays produced in $\bar{p} p $
collisions at $\sqrt{s}=1.96~\TeV$ is reported with a statistical significance 
greater than 5 standard deviations. 
A fit to the $J/\psi\,\phi$ mass spectrum is performed
assuming the presence of a Breit-Wigner resonance. The fit yields a signal of 
$\yieldone$ resonance events, and resonance mass and width of 
 $\massone$ and $\widthone$ respectively.
The parameters of this resonance-like structure are consistent with values 
reported from an earlier CDF analysis.

\end{abstract}


\maketitle


The existence of exotic mesons beyond $q\bar{q}$ has been discussed for 
many years~\cite{PDG}, but evidence for such mesons has not  been clearly 
established. The recent  discoveries of  states   
with charmonium-like decay modes~\cite{x3872discovery,y3940,y4260discovery,y4320} 
that do not fit into 
the overall charmonium system 
have introduced challenges to the conventional $q\bar{q}$ meson model.  
The possible interpretations beyond $q\bar{q}$   
such as hybrid~($q\bar{q}g$) and four-quark states~($q\bar{q}q\bar{q}$)
have revitalized  interest in exotic mesons in the charm 
sector~\cite{conventional,hybrid,molecule,fourquark,estiacharm07,ric}.

Recently, the CDF collaboration has reported evidence for a narrow  
structure near the  $J/\psi\,\phi$ 
threshold  in  $B^+\rightarrow J/\psi\,\phi K^+$ decays produced in $\bar{p} p $ 
collisions at $\sqrt{s}=1.96~\TeV$~\cite{y4140}. 
Charge conjugation is implied throughout this letter. 
Since the mass of this state, termed  $Y(4140)$, is well beyond  
the threshold of open charm-pair production, 
the expected branching fraction into this channel for 
conventional charmonium is tiny. 
The structure is the first observed charmonium-like object decaying 
into a pair of quarkonium states~($c\bar{c}$ and $s\bar{s}$) with a relative 
narrow width,  
a possible signature for an exotic 
meson~\cite{hybrid,ric,hybridfromb,y4140theory}. 
The Belle collaboration has searched for this $J/\psi\,\phi$ structure 
without a firm conclusion 
near the $J/\psi\,\phi$ threshold~\cite{belley4140}.

In this Letter, we report a further study of the structures in the $J/\psi\,\phi$ system 
produced in exclusive $B^+\rightarrow J/\psi\,\phi K^+$ decays 
with $J/\psi\rightarrow \mu^+ \mu^-$ and $\phi\rightarrow K^+K^-$  
reported in Ref.~\cite{y4140}.
This analysis is based on a sample  of $\bar{p} p $ 
collision data  collected by the CDF II detector 
with an integrated luminosity of 6.0 $\ifb$. 
This analysis includes the data used in, and supersedes 
the results of Ref.~\cite{y4140}.

The CDF II detector has been described in detail elsewhere~\cite{cdfii}. 
The important components  for this analysis include the tracking,  
muon, and    time-of-flight~(TOF) systems. 
The tracking system is composed of a  silicon-strip 
vertex detector surrounded by an 
open-cell  drift chamber system 
(COT)
located inside a solenoid with a 1.4 T magnetic field.
The COT and silicon-strip vertex detector are used for the measurement of charged-particle
  trajectories and decay locations.  In addition, the COT provides ionization energy loss
  information, ${dE/dx}$, used for kaon discrimination, while the TOF system provides
  complementary kaon discrimination  information.
The central muon identification system is located radially outside the electromagnetic and hadronic calorimeters
and consists of two sets of drift chambers and scintillation counters.
The central detector covers the pseudorapidity region 
$\left| \eta \right| \le  0.6$ and detects muons 
with $p_T~\ge~1.4~\GeVc$~\cite{eta}, and the outer part 
covers the region  $0.6< \left| \eta \right| <  1.0$ and 
detects muons with $p_T~\ge~2.0~\GeVc$.

In this analysis,
$J/\psi\to\mu^+\mu^-$ events are recorded 
using a dedicated three-level dimuon trigger. The first trigger level 
requires two muon candidates with two COT tracks that extrapolate to 
track segments in the muon detectors.
The second level applies additional  kinematic requirements to the muon pair candidate.
The third level requires the invariant mass of the $\mu^+\mu^-$ pair to be 
within the  mass range of $2.7$  to $4.0$ $\GeVcc$. 
The trigger requirements are confirmed offline.

We apply the same requirements described in 
the previous analysis~\cite{y4140} to the current data. 
We form  $B^+\rightarrow J/\psi\,\phi K^+$ candidates 
by combining a $J/\psi\to\mu^+\mu^-$ candidate, 
a $\phi\rightarrow K^+K^-$ candidate,  and an additional charged track, 
which are consistent with originating from a common point.
The three hadronic tracks must 
be identified as kaon candidates by using a log-likelihood ratio estimator.  
This quantity reflects how well a candidate track can be positively identified 
as a kaon relative to other hadrons with its ${dE/dx}$ and $\mathrm{TOF}$ 
information and must exceed 0.2~\cite{bsmix}.
The reconstructed masses of the $J/\psi$ and $\phi$ meson candidates must 
lie within 50 and 7~\MeVcc~of their nominal values, respectively.
In the final $B^+$ reconstruction the $\mu^+\mu^-$ mass is constrained 
to the known $J/\psi$ mass~\cite{PDG}, 
and the $B^+$ candidates must have $p_T > 4$ \GeVc.
In addition, we require  ${L_{xy}(B^+)} > $~500 $\um$ for the 
$B^+\rightarrow J/\psi\,\phi K^+$ candidate,
where ${L_{xy}(B^+)}$ is the projection onto ${\vec{p}_T(B^+)}$ 
of the vector connecting the primary interaction point, 
determined for each event using  
prompt tracks, to the reconstructed ${B^+}$ decay point.

The invariant mass spectrum of the selected $J/\psi\,\phi K^+$ candidates is 
shown in Fig.~\ref{myfig1}(a). 
It is fit with a Gaussian signal function with its root-mean-square~(RMS) 
width  
fixed to the mass resolution of 5.9 \MeVcc~obtained from Monte Carlo~(MC) 
simulation~\cite{cdfbc} and  mean fixed to the nominal $B^+$  mass~\cite{PDG} 
and a linear background function.
The $B^+$ yield is  $115\pm12(\mathrm{stat})$ events, 
a 53\% increase over the previous  analysis. 
This increase in yield, for an integrated luminosity increased by a 
factor of 2.2, is reduced by a trigger rate limitation at the 
higher instantaneous luminosities for the later data-taking period. 
The yield increase in the complementary  mode 
$B^+\rightarrow J/\psi\, \pi^+\pi^- K^+$ is 51.8$\pm$2.4\%, consistent with 
the yield increase in $B^+ \rightarrow J/\psi\,\phi K^+$ channel.

We then select   $B^+$ signal candidates with a mass 
within  $\pm$3~RMS~($\pm$17.7 \MeVcc) of the nominal $B^+$ mass.
Events with a mass within $\mathrm{[-9,-6]}$~RMS or $\mathrm{[+6,+9]}$~RMS 
of the nominal $B^+$ mass are called $B$ sideband events. 
They are   normalized 
into the $B^+$ signal region assuming a linear background distribution. 
The $J/\psi$ signal, checked by removing its mass constraint, contains almost 
no background. 
Figure~\ref{myfig1}(b) shows the invariant mass distribution 
of the $K^+ K^-$ pairs from $J/\psi\, K^+K^- K^+$ candidates 
inside the $B$ mass window and in the $B$ sidebands before applying the 
restriction on the $K^+K^-$ mass.
The clear $\phi$ signal inside the $B$ mass window and almost featureless 
$K^+K^-$ mass distribution in the $B$ sideband  
indicate that 
the $B^+ \rightarrow J/\psi\, K^+K^-K^+$ final state is well described 
as $J/\psi\, \phi K^+$. 
In none of the candidate events do both $K^+K^-$ combinations from the 
three-kaon final state fall into the $\phi$ mass window.

Fig.~\ref{myfig2} shows 
the mass difference $\Delta M= m(\mu^+\mu^-K^+K^-)- m(\mu^+\mu^-)$  
for events in the $B^+$ mass window.
Events from reference~\cite{y4140} and from new data are shown 
in (a) top and bottom. 
In the $Y(4140)$ signal region ($\Delta M <1.07$\GeVcc), the new data agree within 1$\sigma$ of the 
expectation (6 events compared to 7.3 expected). Over the entire examined region 
the two data sets are consistent at the 7\% probability 
level. We have investigated the consistency of particle ID for the two 
data sets using the  $B^+\rightarrow J/\psi\, K^+$ channel 
and see no discrepant effects. In (b) and (c), we display  $\Delta M$ 
distributions for the events in the $B$ signal and sideband in the 
combined data sample.  
We restrict our study to events with $\Delta M$ smaller than 
1.56 \GeVcc~to avoid appreciable combinatorial backgrounds  from misidentified 
$B^0_s\rightarrow \psi(2S) \phi\rightarrow (J/\psi\,\pi^+\pi^-)\phi$ 
decays~\cite{y4140}.  
An enhancement is observed near the $J/\psi\,\phi$ threshold from the $B^+$ signal 
while there are no events in the $\Delta M$ 
range below 1.1~\GeVcc~from the combinatorial background estimated from $B$ sideband events.

\begin{figure}
\includegraphics[width=9.0cm,height=4.5cm]{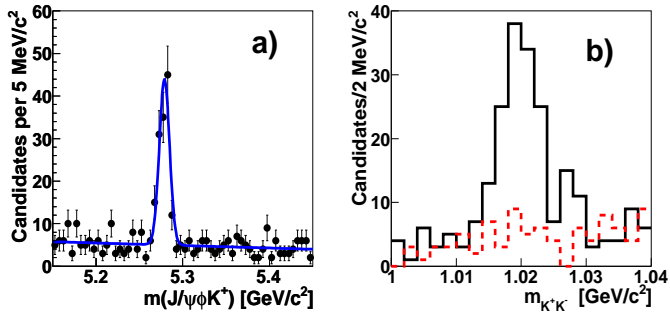}
\caption{\label{myfig1} 
(a) The mass distribution of $J/\psi\,\phi K^+$;
the solid blue line is a fit to the data with a Gaussian signal 
function and linear background function.
(b) The  $K^+K^-$ mass distributions inside the $B$ mass 
window (black solid) and in the $B$ sidebands (red dotted).
}
\end{figure}

\begin{figure}
\includegraphics[width=9.0cm,height=4.5cm]{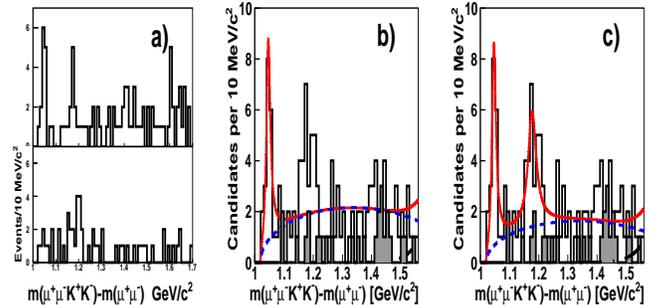}
\caption{\label{myfig2} 
(a) The mass difference, $\Delta M$, between $\mu^+\mu^-K^+K^-$ 
and $\mu^+\mu^-$,   in the $B^+$ mass window. Top--data 
from Ref.~\cite{y4140}, bottom--new data.
(b) A fit to the combined data assuming $Y(4140)$ only.
(c) A fit to the combined data assuming two structures. This fit, including the second peak,  
lowers the 3--body phase space 
background under the first peak and increases its yield and significance 
with negligible effect on its resonance parameters. 
The shaded histogram is the data from the B sideband.
The dotted blue curve is the predicted background contribution, the 
dash-dotted black curve is 
the predicted $B_s^0$ contamination,  and the solid red curve is the total 
unbinned fit. 
}
\end{figure}

We model the observed threshold structure by an $S$-wave relativistic 
Breit-Wigner  (BW) 
function~\cite{sbw} convoluted with 
a Gaussian resolution function with the RMS fixed to 1.7 \MeVcc~obtained from MC.   
Three--body phase space~\cite{PDG} is used to describe the background shape.
There is still a small $B_s^0$ contribution~(3.3$\pm$1.0 events) 
in the $\Delta M$ distribution up to 1.56 GeV. The MC shape of the $B_s^0$ contribution is normalized to this area and added to 
the three-body phase space. 
The parameters from an unbinned likelihood fit to the $\Delta M$ 
distribution, as shown in Fig.~\ref{myfig2}(b), are given  
in Table~\ref{tab:fit1st}.
To test the hypothesis that the structure 
has zero width (weak decay), we also fit the $\Delta M$ distribution 
to a zero-width peak, using a 
single Gaussian with RMS given by the expected mass 
resolution~(1.7 \MeVcc), plus phase space background.
The statistical significance for a non-zero width determined by 
the likelihood ratio between these two fits
is 3.7$\sigma$, favoring a strong decay~(non-zero width) 
rather than a weak decay for this structure.

The combinatorial background contains primarily misidentified $\phi$ 
candidates, as can be seen in Fig.~\ref{myfig1} (b).  
These two tracks with a $\phi$-like mass will be combined
with a real $J/\psi$, and an additional kaon candidate,
all having a common vertex and forming a $B$ mass.
We model this component with phase space.
To check this assumption, we performed 
several studies in which we relaxed cuts that would
not influence the mass-difference distribution of events from the B mass
region: loosened vertex requirements or loosened $L_{xy}$ cuts.  These
studies show that the combinatoric background from the B sideband region
is consistent with 3--body phase space.  
We can now conclude that the flat background hypothesis 
used in the previous paper~\cite{y4140} was overly conservative.

We determine the significance of the structure at the $J/\psi\, \phi$ threshold 
based on simulation.  We generated  
$8.4\times10^7$ mass spectra~(119 events for each, corresponding to 
the number of observed events) drawn from a three--body phase-space-like 
distribution,  and search for the most significant fluctuation in
each spectrum in the mass range of 1.02 to 1.56 \GeVcc, with widths
in the range of resolution up to 120 \MeVcc~\cite{y4140}.
We evaluate  $2\Delta\mathrm{ln}\mathcal{L}={-2{\mathrm{ln}}(\mathcal{L}_0/\mathcal{L}_{{max}})}$ value 
for each generated spectrum,  
where $\mathcal{L}_0$ and $\mathcal{L}_{{max}}$ are the likelihood 
values for the null  hypothesis fit and signal hypothesis fit.
Both fits use three-body phase space to describe the background. 
There are 19 generated spectra with a 
$2\Delta\mathrm{ln}\mathcal{L}$ value 
greater than or equal to the value~(34.9 obtained in the data assuming the $Y(4140)$ structure 
only~\footnote{The $2\Delta\mathrm{ln}\mathcal{L}$ is 40.6 if allowing the existence of the second structure around $\Delta M$ of~1.18~\GeVcc.}) obtained in the data. 
The resulting $p$-value, taken as the fraction of the generated spectra 
with a $2\Delta\mathrm{ln}\mathcal{L}$ value 
greater than or equal to the value obtained in the data, 
is $2.3\times 10^{-7}$, corresponding to a significance greater than 5.0$\sigma$~\footnote{Monte Carlo studies indicate that fits with small numbers of signal
   events tend to return statistical uncertainties that underestimate
   the true uncertainties, an effect that asymptotically decreases
   as the sample size becomes larger. 
Such an underestimate of the fitted parameter uncertainties 
does not influence the evaluation of the signal significance, 
which depends only on the  background fluctuation probability.}.

The mass of this enhancement is $\massfitone$ after adding the 
$J/\psi$ mass~\cite{PDG} to the $\Delta M$ calculation. To study the systematic 
uncertainties of the mass, width, and yield, we repeat the fit to  
the $\Delta M$ distribution 
using a non--relativistic BW and $P$-wave relativistic BW for signal. 
Other systematic uncertainties were also considered, including  
assuming the existence of the second structure, varying the mass resolution and $B_s^0$ component amplitude, as well as the systematic uncertainty due to 
the particle identification. 
The resulting systematic uncertainties for the measured quantities are shown 
in Table~\ref{tab:fit1st}.

\begin{table}[!htb]
\caption{The fit results for $J/\psi\,\phi$ resonance near threshold. 
The first uncertainty is 
statistical, and the second one is systematic.}
\begin{center}
\begin{tabular}{lrr}
\hline
  $\Delta M$ [MeV/c$^2$]&\hspace{0.5cm} Width  [MeV/c$^2$]    &\hspace{1.cm} Yield   \\
\hline\hline
  $\massdifffitone$    &  $\widthfitone$   &  $\yieldonetab$             \\
\hline\hline
\end{tabular}
\label{tab:fit1st}
\end{center}
\end{table}

The relative trigger and reconstruction efficiency  
$\epsilon(B^+\rightarrow Y(4140) K^+)\times \epsilon(Y(4140)\rightarrow J/\psi\,\phi)$/$\epsilon(B^+\rightarrow J/\psi\,\phi K^+)$ is determined to be 1.1, using an $S$-wave 
BW with mean and width values determined from data to represent the $Y(4140)$ 
structure and a three--body phase space kinematics 
for the $B^+\rightarrow J/\psi\,\phi K^+$ decay.
Thus the relative branching fraction   
$\mathcal{B}_{rel}=\mathcal{B}(B^+\rightarrow Y(4140) K^+) \times \mathcal{B}(Y(4140)\rightarrow J/\psi\,\phi)$/$\mathcal{B}(B^+\rightarrow J/\psi\,\phi K^+)$ including systematic uncertainties is $\rebfone$.

An excess above the three-body phase space background shape appears at approximately~1.18~\GeVcc~in Fig.~\ref{myfig2}~(b). 
Since the significance of $Y(4140)$ is greater than 5$\sigma$, we assume 
the existence of the $Y(4140)$ with the parameters given in 
Table~\ref{tab:fit1st} and background given by three--body phase space, 
and we test for the existence of a possible structure around 1.18~\GeVcc~as  
shown in Fig.~\ref{myfig2}~(c). The signal PDF for the second structure is  
an $S$-wave relativistic BW function~\cite{sbw} convoluted with 
a Gaussian resolution function with the RMS fixed to 3.0 \MeVcc~obtained from MC.   
For the second structure 
${-2{\mathrm{ln}}(\mathcal{L}_0/\mathcal{L}_{{max}})}$  
is 16.8, where $\mathcal{L}_0$ and $\mathcal{L}_{{max}}$ are the likelihood 
values for the null  hypothesis fit assuming the $Y(4140)$-only 
and signal hypothesis fit assuming the $Y(4140)$ and a second structure near  
$\Delta M$  $\simeq$~1.18~\GeVcc. The $p$-value determined by a  simulation 
similar to the $Y(4140)$ investigation  
is $1.1\times10^{-3}$, which corresponds to a significance of 3.1$\sigma$.
The fit returns a yield of $22\pm8$ events,   
a $\Delta M$ of $1177.7^{+8.4}_{-6.7}$ \MeVcc, and a width 
of $32.3^{+21.9}_{-15.3}$ \MeVcc~for the structure near $\Delta M$ $\simeq$~1.18~\GeVcc. Refitting the $\Delta M$ distribution with a second structure produces 
negligible changes in the mass and width of the 
$Y(4140)$. The yield of the $Y(4140)$ increases by one event. 
We evaluated the systematic uncertainties for the second structure in the 
same way as for the $Y(4140)$ structure and found 
 systematic uncertainties of 1.9~\MeVcc~for the mass and 7.6~\MeVcc~for the width. The mass of the second structure is  
$\masstwosys$ after including the world-average $J/\psi$ mass.

In summary, the increased $B^+ \rightarrow J/\psi\, \phi K^+$ sample at CDF 
enables us to observe the $Y(4140)$ structure with a significance greater 
than 5$\sigma$. 
Assuming an $S$-wave relativistic BW, 
the mass and width of this structure  
are measured to be $\massone$ 
and $\widthone$, respectively.
They are consistent with the previous 
report~($m=4143.0\pm2.9(\mathrm{stat})\pm1.2 (\mathrm{syst})$~\MeVcc, 
$\Gamma=11.7^{+8.3}_{-5.0}(\mathrm{stat})\pm3.7(\mathrm{syst})$~\MeVcc~\cite{y4140}. 
The relative branching fraction 
is determined to be 
$\mathcal{B}_{rel}$ = $\rebfone$. 
Light meson vector-vector threshold enhancements have been seen~\cite{lightmeson}. 
We do not know of any non-exotic mechanism for producing a threshold
enhancement involving a pair of heavy quarkonium states, 
but we cannot exclude the possibility.
We also find evidence for a  second structure with a  
mass of $\masstwosys$, 
a width of $\widthtwosys$ and a yield of $\yieldtwo$ events.
The significance of the second  structure is estimated to be 
approximately 3.1$\sigma$.

\input acknow_PRL.tex


\end{document}

%% file: March2010_Authors.tex
\affiliation{Institute of Physics, Academia Sinica, Taipei, Taiwan 11529, Republic of China} 
\affiliation{Argonne National Laboratory, Argonne, Illinois 60439, USA} 
\affiliation{University of Athens, 157 71 Athens, Greece} 
\affiliation{Institut de Fisica d'Altes Energies, ICREA, Universitat Autonoma de Barcelona, E-08193, Bellaterra (Barcelona), Spain} 
\affiliation{Baylor University, Waco, Texas 76798, USA} 
\affiliation{Istituto Nazionale di Fisica Nucleare Bologna, $^z$University of Bologna, I-40127 Bologna, Italy} 
\affiliation{University of California, Davis, Davis, California 95616, USA} 
\affiliation{University of California, Los Angeles, Los Angeles, California 90024, USA} 
\affiliation{Instituto de Fisica de Cantabria, CSIC-University of Cantabria, 39005 Santander, Spain} 
\affiliation{Carnegie Mellon University, Pittsburgh, Pennsylvania 15213, USA} 
\affiliation{Enrico Fermi Institute, University of Chicago, Chicago, Illinois 60637, USA}
\affiliation{Comenius University, 842 48 Bratislava, Slovakia; Institute of Experimental Physics, 040 01 Kosice, Slovakia} 
\affiliation{Joint Institute for Nuclear Research, RU-141980 Dubna, Russia} 
\affiliation{Duke University, Durham, North Carolina 27708, USA} 
\affiliation{Fermi National Accelerator Laboratory, Batavia, Illinois 60510, USA} 
\affiliation{University of Florida, Gainesville, Florida 32611, USA} 
\affiliation{Laboratori Nazionali di Frascati, Istituto Nazionale di Fisica Nucleare, I-00044 Frascati, Italy} 
\affiliation{University of Geneva, CH-1211 Geneva 4, Switzerland} 
\affiliation{Glasgow University, Glasgow G12 8QQ, United Kingdom} 
\affiliation{Harvard University, Cambridge, Massachusetts 02138, USA} 
\affiliation{Division of High Energy Physics, Department of Physics, University of Helsinki and Helsinki Institute of Physics, FIN-00014, Helsinki, Finland} 
\affiliation{University of Illinois, Urbana, Illinois 61801, USA} 
\affiliation{The Johns Hopkins University, Baltimore, Maryland 21218, USA} 
\affiliation{Institut f\"{u}r Experimentelle Kernphysik, Karlsruhe Institute of Technology, D-76131 Karlsruhe, Germany} 
\affiliation{Center for High Energy Physics: Kyungpook National University, Daegu 702-701, Korea; Seoul National University, Seoul 151-742, Korea; Sungkyunkwan University, Suwon 440-746, Korea; Korea Institute of Science and Technology Information, Daejeon 305-806, Korea; Chonnam National University, Gwangju 500-757, Korea; Chonbuk National University, Jeonju 561-756, Korea} 
\affiliation{Ernest Orlando Lawrence Berkeley National Laboratory, Berkeley, California 94720, USA} 
\affiliation{University of Liverpool, Liverpool L69 7ZE, United Kingdom} 
\affiliation{University College London, London WC1E 6BT, United Kingdom} 
\affiliation{Centro de Investigaciones Energeticas Medioambientales y Tecnologicas, E-28040 Madrid, Spain} 
\affiliation{Massachusetts Institute of Technology, Cambridge, Massachusetts 02139, USA} 
\affiliation{Institute of Particle Physics: McGill University, Montr\'{e}al, Qu\'{e}bec, Canada H3A~2T8; Simon Fraser University, Burnaby, British Columbia, Canada V5A~1S6; University of Toronto, Toronto, Ontario, Canada M5S~1A7; and TRIUMF, Vancouver, British Columbia, Canada V6T~2A3} 
\affiliation{University of Michigan, Ann Arbor, Michigan 48109, USA} 
\affiliation{Michigan State University, East Lansing, Michigan 48824, USA}
\affiliation{Institution for Theoretical and Experimental Physics, ITEP, Moscow 117259, Russia}
\affiliation{University of New Mexico, Albuquerque, New Mexico 87131, USA} 
\affiliation{Northwestern University, Evanston, Illinois 60208, USA} 
\affiliation{The Ohio State University, Columbus, Ohio 43210, USA} 
\affiliation{Okayama University, Okayama 700-8530, Japan} 
\affiliation{Osaka City University, Osaka 588, Japan} 
\affiliation{University of Oxford, Oxford OX1 3RH, United Kingdom} 
\affiliation{Istituto Nazionale di Fisica Nucleare, Sezione di Padova-Trento, $^{aa}$University of Padova, I-35131 Padova, Italy} 
\affiliation{LPNHE, Universite Pierre et Marie Curie/IN2P3-CNRS, UMR7585, Paris, F-75252 France} 
\affiliation{University of Pennsylvania, Philadelphia, Pennsylvania 19104, USA}
\affiliation{Istituto Nazionale di Fisica Nucleare Pisa, $^{bb}$University of Pisa, $^{cc}$University of Siena and $^{dd}$Scuola Normale Superiore, I-56127 Pisa, Italy} 
\affiliation{University of Pittsburgh, Pittsburgh, Pennsylvania 15260, USA} 
\affiliation{Purdue University, West Lafayette, Indiana 47907, USA} 
\affiliation{University of Rochester, Rochester, New York 14627, USA} 
\affiliation{The Rockefeller University, New York, New York 10065, USA} 
\affiliation{Istituto Nazionale di Fisica Nucleare, Sezione di Roma 1, $^{ee}$Sapienza Universit\`{a} di Roma, I-00185 Roma, Italy} 

\affiliation{Rutgers University, Piscataway, New Jersey 08855, USA} 
\affiliation{Texas A\&M University, College Station, Texas 77843, USA} 
\affiliation{Istituto Nazionale di Fisica Nucleare Trieste/Udine, I-34100 Trieste, $^{ff}$University of Trieste/Udine, I-33100 Udine, Italy} 
\affiliation{University of Tsukuba, Tsukuba, Ibaraki 305, Japan} 
\affiliation{Tufts University, Medford, Massachusetts 02155, USA} 
\affiliation{University of Virginia, Charlottesville, VA  22906, USA}
\affiliation{Waseda University, Tokyo 169, Japan} 
\affiliation{Wayne State University, Detroit, Michigan 48201, USA} 
\affiliation{University of Wisconsin, Madison, Wisconsin 53706, USA} 
\affiliation{Yale University, New Haven, Connecticut 06520, USA} 
\author{T.~Aaltonen}
\affiliation{Division of High Energy Physics, Department of Physics, University of Helsinki and Helsinki Institute of Physics, FIN-00014, Helsinki, Finland}
\author{B.~\'{A}lvarez~Gonz\'{a}lez$^v$}
\affiliation{Instituto de Fisica de Cantabria, CSIC-University of Cantabria, 39005 Santander, Spain}
\author{S.~Amerio}
\affiliation{Istituto Nazionale di Fisica Nucleare, Sezione di Padova-Trento, $^{aa}$University of Padova, I-35131 Padova, Italy} 

\author{D.~Amidei}
\affiliation{University of Michigan, Ann Arbor, Michigan 48109, USA}
\author{A.~Anastassov}
\affiliation{Northwestern University, Evanston, Illinois 60208, USA}
\author{A.~Annovi}
\affiliation{Laboratori Nazionali di Frascati, Istituto Nazionale di Fisica Nucleare, I-00044 Frascati, Italy}
\author{J.~Antos}
\affiliation{Comenius University, 842 48 Bratislava, Slovakia; Institute of Experimental Physics, 040 01 Kosice, Slovakia}
\author{G.~Apollinari}
\affiliation{Fermi National Accelerator Laboratory, Batavia, Illinois 60510, USA}
\author{J.A.~Appel}
\affiliation{Fermi National Accelerator Laboratory, Batavia, Illinois 60510, USA}
\author{A.~Apresyan}
\affiliation{Purdue University, West Lafayette, Indiana 47907, USA}
% ...
\collaboration{CDF Collaboration\footnote{With visitors from $^a$University of Massachusetts Amherst, Amherst, Massachusetts 01003,
$^b$Istituto Nazionale di Fisica Nucleare, Sezione di Cagliari, 09042 Monserrato (Cagliari), Italy,
$^c$University of California Irvine, Irvine, CA  92697, 
$^d$University of California Santa Barbara, Santa Barbara, CA 93106
$^e$University of California Santa Cruz, Santa Cruz, CA  95064,
$^f$CERN,CH-1211 Geneva, Switzerland,
$^g$Cornell University, Ithaca, NY  14853, 
$^h$University of Cyprus, Nicosia CY-1678, Cyprus, 
$^i$University College Dublin, Dublin 4, Ireland,
$^j$University of Fukui, Fukui City, Fukui Prefecture, Japan 910-0017,
$^k$Universidad Iberoamericana, Mexico D.F., Mexico,
$^l$Iowa State University, Ames, IA  50011,
$^m$University of Iowa, Iowa City, IA  52242,
$^n$Kinki University, Higashi-Osaka City, Japan 577-8502,
$^o$Kansas State University, Manhattan, KS 66506,
$^p$University of Manchester, Manchester M13 9PL, England,
$^q$Queen Mary, University of London, London, E1 4NS, England,
$^r$Muons, Inc., Batavia, IL 60510,
$^s$Nagasaki Institute of Applied Science, Nagasaki, Japan, 
$^t$National Research Nuclear University, Moscow, Russia,
$^u$University of Notre Dame, Notre Dame, IN 46556,
$^v$Universidad de Oviedo, E-33007 Oviedo, Spain, 
$^w$Texas Tech University, Lubbock, TX  79609, 
$^x$Universidad Tecnica Federico Santa Maria, 110v Valparaiso, Chile,
$^y$Yarmouk University, Irbid 211-63, Jordan,
$^{gg}$On leave from J.~Stefan Institute, Ljubljana, Slovenia, 
}}
\noaffiliation

%% file: acknow_PRL.tex
\par 
We thank the Fermilab staff and the technical staffs of the participating institutions for their vital 
contributions. We wish to thank E. Eichten, S. Brodsky and C. Quigg for helpful discussions. 
This work was supported by the U.S. Department of Energy and National Science Foundation; the 
Italian Istituto Nazionale di Fisica Nucleare; the Ministry of Education, Culture, Sports, Science and Technology 
of Japan; the Natural Sciences and Engineering Research Council of Canada; the National Science Council of the 
Republic of China; the Swiss National Science Foundation; the A.P. Sloan Foundation; the Bundesministerium f\"ur 
Bildung und Forschung, Germany; the Korean Science and Engineering Foundation and the Korean Research Foundation; 
the Science and Technology Facilities Council and the Royal Society, UK; the Institut National de Physique 
Nucleaire et Physique des Particules/CNRS; the Russian Foundation for Basic Research; the Ministerio de Ciencia e 
Innovaci\'{o}n, and Programa Consolider-Ingenio 2010, Spain; the Slovak R\&D Agency; and the Academy of Finland.

%% file: paper.bbl
\begin{thebibliography}{99}
\bibitem{eta}
We use a coordinate system in which the proton beam
      direction is defined as the $z$ axis.  
The angle $\theta$ is the usual
   polar angle.  We define the
   pseudorapidity as  $\eta\equiv -\ln(\tan\frac{\theta}{2})$.  
   The transverse momentum is defined as  $p_T =p\sin\theta$, where $p$ is the momentum measured in 
   the tracking system.
\bibitem{sbw}
$\frac{dN}{dm} \propto \frac{m \Gamma(m)}{(m^2-m_0^2)^2+m_0^2\Gamma^2(m) }$, where
$\Gamma(m)=\Gamma_0\frac{q}{q_0}\frac{m_0}{m}$, $q$ is the daughter momentum in the mother rest frame, and the 0 subscript indicates the 
value at the pole mass.
\end{thebibliography}
